\documentclass[usenatbib]{mn2e}
\usepackage{times}
\usepackage{epsfig}

\newcommand{\eg}{e.$.\!$g.\ }
\newcommand{\be}{\begin{equation}}
\newcommand{\ee}{\end{equation}}
\newcommand{\etal}{et al.\ }
\newcommand{\Chandra}{{\it Chandra} }
\newcommand{\XMM}{{\it XMM-Newton} }
\newcommand{\CIAO}{{\it CIAO} }

\newcommand{\ki}{$\chi^2$\,}

\title[Chandra observations of a nearby fossil group]
{An old galaxy group: Chandra X-ray observations of the nearby
  fossil group NGC 6482}
\author[H. G. Khosroshahi, L. R. Jones and T. J. Ponman]{Habib G. Khosroshahi\thanks{E-mail:
habib@star.sr.bham.ac.uk (HGK); tjp@star.sr.bham.ac.uk (TJP);
lrj@star.sr.bham.ac.uk (LRJ)},
Laurence R. Jones \& Trevor J. Ponman\\
School of Physics and Astronomy, The University of Birmingham,
Birmingham B15 2TT, UK}

\begin{document}

\date{Accepted, Received}

\pagerange{\pageref{firstpage}--\pageref{lastpage}} \pubyear{2003}

\maketitle

\label{firstpage}

\begin{abstract} 
We present the first detailed X-ray observations, using \Chandra, of
NGC\,6482 -- the nearest known `fossil group'. The group is dominated
by an optically luminous giant elliptical galaxy and all other known
group members are at least two magnitudes fainter.  The global X-ray
properties (luminosity, temperature, extent) of NGC\,6482 fall within
the range of other groups, but the detailed properties show
interesting differences.  We derive the gas temperature and total mass
profiles for the central 30\,$h_{70}^{-1}$ kpc ($\sim 0.1 r_{200}$)
using ACIS spatially resolved spectroscopy.  The unusually high
$L_X/L_{opt}$ ratio is found to result from a high central gas
density.  The temperature profile shows a continuous decrease outward,
dropping to 0.63 of its central value at 0.1$r_{200}$.  The derived
total mass profile is strongly centrally peaked, suggesting an early
formation epoch. These results support a picture in which fossil
groups are old, giving time for the most massive galaxies to have merged
(via the effects of dynamical friction) to produce a central giant
elliptical galaxy.

Although the cooling time within $0.1 r_{200}$ is less than a Hubble time,
no decrease in central temperature is detected.  The entropy of the
system lies toward the low side of the distribution seen in poor
groups, and it drops all the way into the centre of the system,
reaching very low values. No isentropic core, such as those predicted
in simple preheating models, is present. Given the lack of any central
temperature drop in the system, it seems unlikely that radiative cooling
can be invoked to explain this low central entropy. The lack of any
signature of central cooling is especially striking in a system
which appears to be old and relaxed, and to have a central cooling
time $\le 10^8$ years. We find that the centrally peaked temperature
profile is consistent with a steady-state cooling flow solution
with an accretion rate of 2~M$_{\odot}$ yr$^{-1}$, given
the large $P\,dV$ work arising from the cuspy mass profile. However,
solutions involving distributed or non-steady heating cannot be ruled
out.
\end{abstract}


\section{Introduction}

In a cosmological hierarchy where smaller systems form prior to the
collapse and virialization of more massive systems, groups form before
clusters.  However, distinguishing old groups among all the groups
observed today has generally proved impossible.  In galaxy groups,
galaxy mergers can occur efficiently because of their low velocity
dispersion, similar in some cases to the internal velocity dispersions
of galaxies.  In an old, relatively isolated group with little
subsequent infall, there may be sufficient time for most massive
galaxies (but not the low mass galaxies) to lose energy via dynamical
friction and merge, producing a system consisting of a giant central
elliptical galaxy, dwarf galaxies and an extended X-ray group halo.
This is one of the proposed scenarios for the formation of giant
isolated elliptical galaxies and has its origins in the galactic
cannibalism of \citet{ho78}.  The observation of `fossil' systems
\citep{ponman94} gave the first direct clue to the above
mechanism. Numerical simulations \citep{barnes89} suggest that a few
billion years is required for compact group members to merge and form
a single elliptical galaxy. It is then feasible to form such a system
in a Hubble time, but if it falls into a cluster during this process
then the evolutionary history will be obscured.

There are about a dozen such systems identified \citep{ponman94,
vikh99, mz99, jones00, romer00, mat01, jones03} but
few of them have been studied in detail. Based on their space density,
and using a precise definition of such a system,
\citet{jones03} estimate that fossil systems represent 8\%-20\% of
all systems with the same X-ray luminosity, and are as numerous as
poor and rich clusters combined. Therefore studying these systems
gives us information about the formation of giant elliptical galaxies
at redshifts which are readily accessible and without the additional
complications of infall into clusters.

\citet{jones03} found that fossil systems had an unusually high
$L_X/L_{opt}$ ratio compared to normal groups, and suggested that the
high X-ray luminosity was due to a high density of hot gas, reflecting
an early epoch of formation when the Universe was denser.
Alternatively, \citet{vikh99} suggested that such systems have
an unusually high mass-to-light ratio and thus represent partially
failed groups with low star formation efficiencies.  If the
distinctive properties of fossil groups arise from an unusually early
collapse epoch, then they should have strongly peaked mass profiles
(\eg Navarro \etal 1995).  They are expected to contain central cool
gas, because no merger induced mixing has taken place recently. It has
also been suggested that these systems should contain lower gas entropy
\citep{jones03}. A given amount of energy injected at a high
density epoch should result in lower entropy than if it is injected at
a later, low density, epoch.

The discovery of fossils, and measurements of their X-ray properties,
has so far been based on ROSAT observations. In this paper we report
on a detailed \Chandra observation of the nearest known fossil group.

This paper is organised as following: Section 2 briefly reviews the
group properties and describes the data and preparations. The results
from imaging and spectral X-ray analysis are presented in section
3. Section 4 describes the distribution of mass and mass-to-light
ratio. X-ray scaling relations are discussed in section 5. A
discussion and concluding remarks are summarised in section 6.

\section{Observation and Preparation}

\subsection{The group}

The NGC\,6482 group was selected for \Chandra observations via a search
for nearby fossil groups, based on a sample of isolated optically
luminous elliptical galaxies. The isolation criteria were that all
galaxies listed in NED within 0.75 Mpc and 1000 km s$^{-1}$ of the
elliptical galaxies were required to be more than 2 magnitudes
fainter. In addition, the X-ray luminosity, based on the ROSAT All-Sky
Survey, was required to be $>$10$^{42}$ erg s$^{-1}$, much greater
than normal ellipticals and meeting the \citet{jones03} definition of
a fossil group.

The closest galaxy [or group] found which satisfied these criteria was 
NGC\,6482 (RA. 17:51:48.83 and
Dec. +23:04:18.9; J2000), at a redshift $z=0.0131$. This is a giant
elliptical galaxy and brightest group member. There are 5 confirmed
members of this group within $\sim 60$ arcmin of the brightest member.
These are NGC\,6482, UGC 11018, MRK 895, UGC 11024 and CGCG
141-014. The second brightest galaxy, UGC 11024, has a total (2MASS) J
magnitude of 11.41, compared with J$_{tot}$=9.35 for NGC\,6482. Thus
all the known members are indeed more than 2 magnitudes fainter than
the brightest galaxy.

There are more than 20 2MASS near-IR galaxy detections within $\sim
20$ arcmin, the area limited to the locations of the above members,
with a total J-magnitude ranging from 14.2 to 15.5, but there is no
independent source to confirm their group membership. The velocity
dispersion of the group is $\approx$242 km s$^{-1}$ calculated using
the heliocentric velocities of individual group members (NED, NASA/IPAC Extragalactic Database).

Analysis of a pointed ROSAT PSPC observation showed that NGC\,6482 was
at the centre of extended X-ray emission. The X-ray luminosity was
$L_{X,bol}=1.3\times10^{42}$ erg s$^{-1}$.  An X-ray point source
coincident with NGC\,6482, probably an AGN, hampered a detailed
analysis. \citet{gou94} have noted the presence of a radio source and 
emission-line gas and dust in the nucleus of NGC\,6482, probably also
associated with the active nucleus. The derived mass of ionised gas
and dust is typical of other elliptical galaxies in the study.  This
study also shows that there is no strong dust lane in the brightest
group galaxy suggesting the absence of any recent merger induced
mixing.

$H_0=70$ km~s$^{-1}$~Mpc$^{-1}$ and $\Omega_m=0.3$ are assumed throughout 
this paper. At the redshift of NGC\,6482 the luminosity and diameter
distances are 56.7 Mpc and 55.2 Mpc, respectively and 1 arcsec $\equiv$ 
0.268 kpc.

\begin{figure}
\center
\epsfig{file=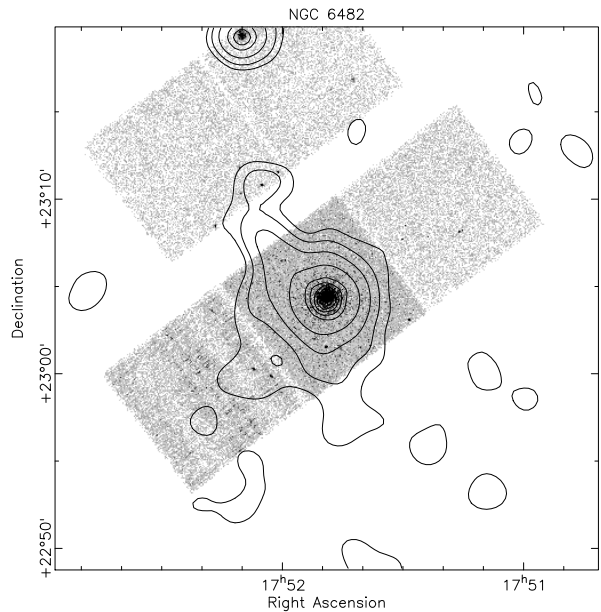,width=3.2in}
\caption{ACIS raw image of NGC\,6482, centred on S3, in the full energy 
range. X-ray contours from ROSAT observation is overlaid.}
\label{fig1}
\end{figure}

\subsection{\Chandra observation and data preparation}

NGC\,6482 was observed using \Chandra ACIS-S on 2002 May 20.
The target was located at the centre of chip 7 (S3), a
back-illuminated CCD.  The total
exposure time was over 19 ks. Bad pixels were removed using the
supplied bad pixel file. Background flares were also removed from
the event 2 file leaving us with a useful exposure time of just
over 18 ks. This study is limited to the S3 chip which is entirely
illuminated by the X-ray emission from the source. A fraction of
the X-ray emission is lost in the gaps between the chips (Fig.
\ref{fig1}) and closest part of chip 2 (to the NE) contains point sources
which appear to be the main source of the observed elongation in the ROSAT
image. For this analysis, unless stated otherwise, the standard CIAO v2.3
routines and tools were used.

\subsubsection{Spatial analysis}

Based on our ROSAT analysis, it was suspected that the X-ray emission
extends beyond the chip under study, S3, and that blank sky
observations must be used to estimate the background.  
We use those provided in the {\it CAIO} calibration data base {\it CALDB}.
The lowest flux regions at the edges of the S3, which are the closest
to being `source free', show a 30\% higher soft X-ray flux on
average compared to the flux from the same regions at the re-projected
blank sky observation. At the same time the hard X-ray count rate
measured for the same regions shows no offset, confirming that the
diffuse source emission has fully covered the chip. The
background estimated from surrounding chips cannot be directly used
since they are all front-illuminated. Thus a background image from
blank sky observations with energies from 0.3-2.0 
keV was used for the spatial analysis.

Point sources were detected using the CIAO wavelet detection algorithm
and replaced by their local surrounding mean counts before any spatial
analysis was carried out. Vignetting and other sensitivity variations
were corrected for, using an exposure map appropriate for the energy
band.

\begin{figure}
\center \epsfig{file=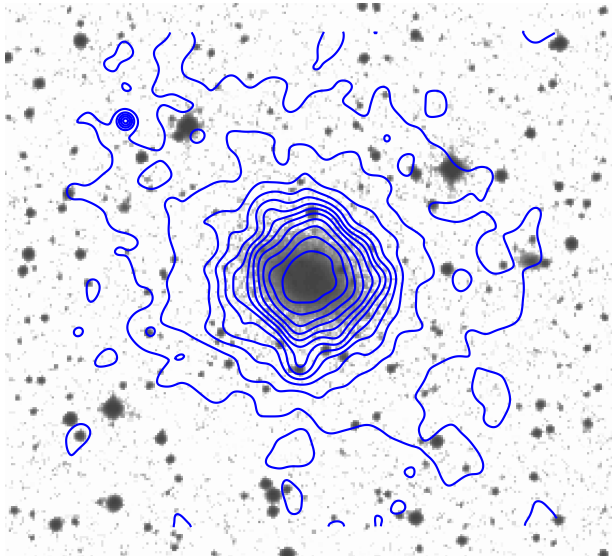,width=3.2in} 
\caption{X-ray contours from the soft (0.3 - 1 keV) diffuse emission 
overlaid on a $8\times 8$ arcmin size DSS optical image.} 
\label{fig2}
\end{figure}

\subsubsection{Spectral analysis}

We extracted ACIS spectra in successive circular annuli, in the energy
range limited to 0.5-2.0 keV, excluding point sources. The background
was chosen for each annulus separately, to account for vignetting, and
from the same region on the re-projected blank sky observations to
account for local variations within the chip.

The spectra were fit with absorbed hot plasma models and the APEC 
\citep{smith01} model was found to give the best fit.  A fixed 
hydrogen column density
of $N_{H, gal}=0.08\times 10^{22}\, {\rm cm}^{-2}$ was included in the
model to account for Galactic absorption.  An integrated spectrum was
first extracted from a large region, of 3 arcmin radius.  A mean
temperature of 0.66 keV and an unabsorbed flux of
$0.175\times10^{-12}$ erg s$^{-1}$ cm$^{-2}$ (0.5-2 keV) was found.
However, the fit was poor with a reduced \ki of about 1.5. This high
\ki from a single temperature fit is not surprising, since we will see
below that a strong temperature gradient is present in the X-ray
emitting plasma. Absorption intrinsic to the source was also required
by the fit, with a value of $0.14(\pm0.02)\times 10^{22}\, {\rm
cm}^{-2}$.

\section{Results}

\subsection{X-ray morphology and surface brightness}

The diffuse 0.3-2 keV X-ray emission, (see Fig \ref{fig2}), shows a rather
relaxed X-ray morphology in the core of the group. For the production of
this image, point sources were removed, the background subtracted,
and the $8\times8$ binned image adaptively smoothed and corrected using
the exposure map.  The X-ray contours in the central region are
circularly symmetric but the outer region shows a moderate NE-SW
elongation.

\subsubsection{Hardness ratio map}

\begin{figure}
\center
\epsfig{file=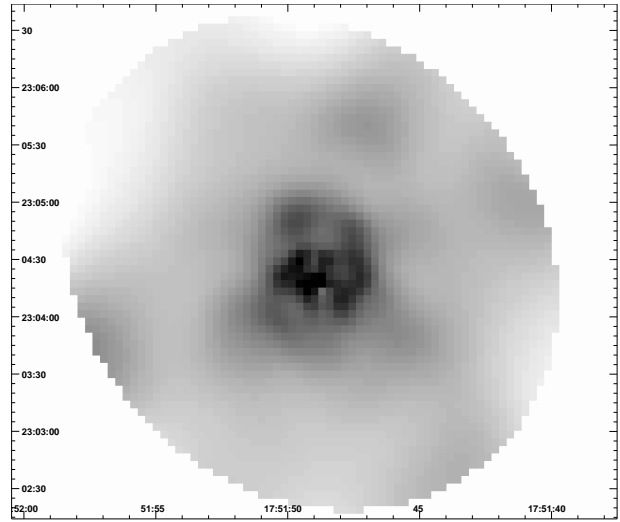,width=3.2in}
\caption{Hardness ratio map of NGC\,6482 on a linear scale. The dark 
region at the centre is the hardest region and the white 
outer region the softest.}
\label{fig3}
\end{figure}

To study any possible substructure and its significance, a
hardness ratio map was obtained by dividing the adaptively smoothed
diffuse emission image in a hard band by that in a soft band. The
soft image contains photons with energies from 400 eV to
900 eV. The hard image consisted of photons with energy ranging from
900 eV to 2 keV. We replaced the point sources by their surrounding
counts before smoothing the hard band $8\times8$ binned image
using the $csmooth$ task in \CIAO. The same scale map is then used to
smooth the soft band image as well as the associated background
and exposure map images (i.e. all these are smoothed in an identical 
way). Before deriving the
hardness ratio map, the corresponding backgrounds were subtracted from
the soft and hard images.  The hardness ratio map is limited to
the central 3 arcmin radius, Fig. \ref{fig3}, due to the limited
S/N. The relatively uniform and circularly symmetric distribution
of the hardness ratio suggests that the system is relatively
relaxed and has not experienced a violent or strong perturbation
recently,  as caused by mergers and starforming
activity. Small scale features at the centre
of the map are found to be at the noise level and most likely due
to the unresolved point sources or inaccurate replacement by local
diffuse emission after point source removal . We therefore assume
that the system is spherically symmetric and is in hydrostatic
equilibrium.

\subsubsection{Gas density profile}
The radial surface brightness profile extracted from the inner 200
arcsec of the 0.3-2.0~keV image, was fitted using a one-dimensional
$\beta$-model,
\be
\Sigma(r)=\Sigma_0[1+(\frac{r}{r_0})^2]^{-3\beta+1/2},
\ee
where $r_0$ and $\beta$ are the core radius and index, and
$\Sigma_0$ the central surface brightness.

Under the assumptions of spherical symmetry and isothermal gas, the above 
surface brightness profile implies a 3-dimensional gas number density density
distribution given by
\be
n(r)=n_0[1+(\frac{r}{r_0})^2]^{-3\beta/2}.
\ee

Fitting the $\beta$-model given by equation 1, excluding the central 
point source, we found $\beta=0.53$
and core radius, $r_0= 7.5$ arcsec, but the reduced \ki was large,
7.5. Excluding 5 arcsec from the central region improved the \ki
by factor of 2 but it is still far from a good fit to the surface
brightness at the central region of the system which is crucial
in the estimation of other physical quantities such as mass and gas
density.

\begin{figure}
\center
\epsfig{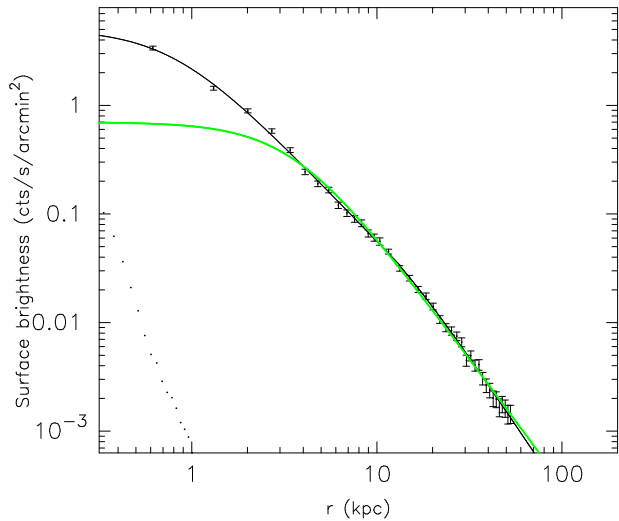}
\caption{Surface brightness profile of NGC\,6482. The solid line represents
the one-dimensional double $\beta$-model excluding the central point 
source. The gray profile is the best single
$\beta$-model fit when the central 5 arcsec (1.34 kpc) is masked out. Dotted 
profile shows the observed PSF profile normalised to the surface brightness 
profile at the centre.}
\label{fig4}
\end{figure}

Thus we used a double $\beta$ model where an isothermal $\beta$-model is
fit separately to the inner and outer regions of the radial
surface brightness distribution The surface brightness in this case is
given by

\be
\begin{array}{ll}
\Sigma_1(r)=\Sigma_{01}[1+(\frac{r}{r_{_{01}}})^2]^{-3\beta_1+1/2},& r\le 
R_{cut}\\
&\\
\Sigma_2(r)=\Sigma_{02}[1+(\frac{r}{r_{_{02}}})^2]^{-3\beta_2+1/2},& r>R_{cut}
\end{array}
\ee 
The surface brightness, and its slope, from
the above equations must be continuous at the cut radius,
$R_{cut}$, which itself is chosen to minimise \ki. In
addition, the continuity of the gas density derivative must be met
in order to derive a unique mass at $R_{cut}$. These constraints 
\citep{parnud02} were directly applied while fitting the model in {\it Sherpa}.

The results from this fit are $r_{01}=2.95\pm0.027,\, r_{02}=26.99\pm0.22$ 
arcsec and
$\beta_1=0.46\pm0.002,\, \beta_2=0.59\pm0.004$ with $R_{cut}=40$ arcsec. 
For comparison, a ROSAT study of the system \citep{sand03} gave a
value $\beta$=0.48, and a poorly constrained core radius of $\sim$13
arcsec.  Fig. \ref{fig4} shows the radial fit to the surface
brightness profile from the double $\beta$-model excluding the central
point source. The gray profile represents the single $\beta$-model fit
to the surface brightness profile when the central 5 arcsec is
excluded. The obvious improvement in the fit is also quantified by the
reduced $\chi^2$ of 0.9.

The central gas density is required to fully describe the distribution
of the gas distribution. This can be obtained from spectral
analysis of a reasonably large region around the centre. The method is
described in the following section and further by \citet{ben03}.

\begin{figure}
\center
\epsfig{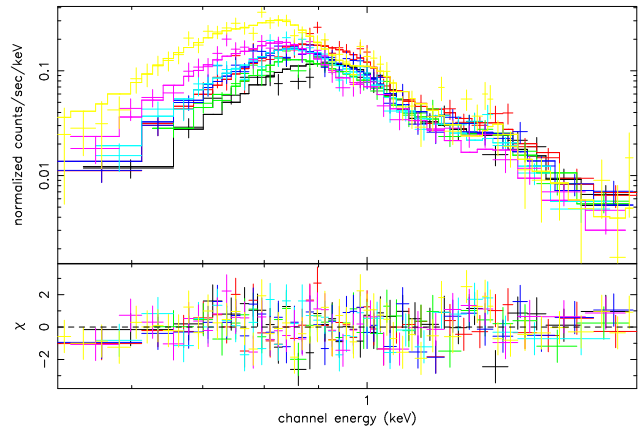}
\caption{Deprojected APEC model fit to the spectra from 7 successive 
annuli extending to a radius of
150 arcsec. The total reduced \ki is 1.01.}
\label{fig5}
\end{figure}

\subsection{Spatially resolved spectroscopy}

An absorbed hot plasma APEC model was fit to spectra extracted from 7
annuli extending to a radius of 150 arcsec. The annuli were chosen to contain
1000 to 1100 net counts, excluding point sources. The spectrum of each
annulus was rebinned to contain a minimum of 20 counts per bin to
increase the S/N. The Ancillary Response File (ARF), representing
effective area as a function of photon energy, was calculated for each
annulus since it varies across the chip.  Due to the extra absorption
introduced to the low energy band, because of the hydrocarbons built
up on the CCDs since the telescope launch, an additional correction
was needed.  This correction was applied to each individual ARF. The
Redistribution Matrix File (RMF), which is used to map the photon
energy to pulse height, was also obtained for each annulus. Since the
X-ray emission covers the entire chip, spectra of the background
were extracted for each annulus from the same region of the blank sky
observations.

The absorbed APEC model was fit to the extracted spectra from 7
annuli. APEC gave a better fit to the data compared to other hot
plasma models such as MEKAL. The innermost region fitted with a relatively
high reduced \ki, and hence more complex models were investigated, as
discussed in Section 3.2.2. With the current data we are not able to
constrain the abundances even when a large region is chosen. Thus we
assume a similar abundance for all the annuli resulting in a value of
Z=$0.76\pm0.28$ Z$_{\odot}$ in XSPEC's default, ANGR, system
\citep{andres89}. Intrinsic absorption was again required in the fits.

To correct for the effect of projection, de-projection was performed
in XSPEC using PROJCT model.  The reduced \ki from the fit to the 7
spectra simultaneously was found to be 1.01 with 248 degrees of
freedom. The PROJCT model fit to the spectra from all annuli is shown
in Fig.~\ref{fig5}.  Bolometric luminosities and intrinsic absorption
of the shells are presented in Table 1. The intrinsic absorption is
consistent with zero in the outermost annuli, as expected if it is
associated with the galaxy. 

\begin{figure}
\center \epsfig{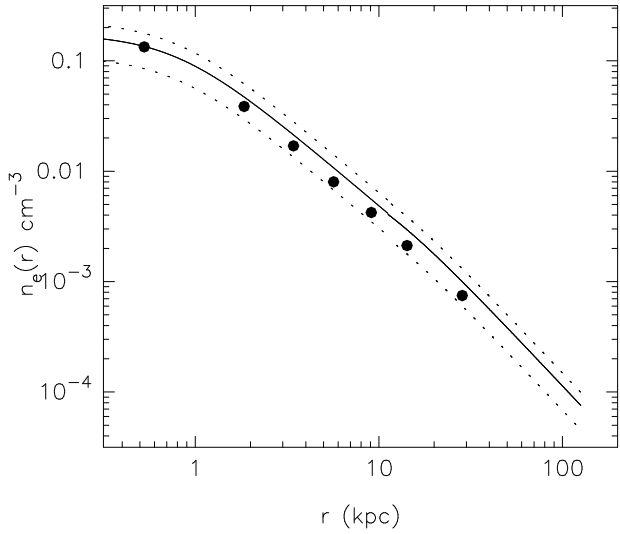}
\caption{Electron number density profile. The data points are the
avraged electron density for each shell measured directly from the
X-ray emission (section 3.3). The solid curve shows the electron number 
density obtained from the double $\beta$-model fit.  Dottted curves 
correspond to the error in the central gas density (see text).} 
\label{fig6}
\end{figure}

In order to derive the central gas density, a circular
region with a radius of $r_1=40$ arcsec was chosen and an APEC model was
fit to the extracted spectrum. The APEC normalisation is defined as:
\be
N_{Apec}=\frac{10^{-14}}{4\pi D_a^2(1+z)^2}\int n_e n_H dV~~cm^{-5}
\ee
where $D_a$ is the angular distance to the source. With the
$\beta$-model fitted to the inner region we calculate the integral
which is now written as the following:

\be
\int n_e n_H dV=1.17\int_0^{\infty}n_H^2(r) 4\pi r^2 (1-\cos \theta)dr.
\ee
where $\tan\theta=r_1/r$ and assuming that $n_e=1.17 n_H$. Knowing
the APEC normalisation from the spectral fit, we found 
the central hydrogen nummber density, $n_{H_{01}}=0.151\pm 0.061$ cm$^{-3}$. 
The value of $n_{H_{02}}=0.010$ cm$^{-3}$ (see equations 3) was then obtained 
from the continuity of the density at $R_{cut}=40$ arcsec.

The above calculations are based on the assumption that the emissivity
of the gas simply scales as density squared. Given the steep
temperature gradient, this is not obviously justified. However, at
these temperatures, emissivity is not a strong function of
temperature.  In order to estimate the effect of temperature
variations on the value of electron density, we calculate the electron
number density directly from the emission of each shell (section 3.3)
and compare with the model 3D electron number density obtained from
the double $\beta$-model fit. Fig \ref{fig6} shows that the measured
electron density is in reasonable agreement with that obtained from
the $\beta$-model analysis.

\begin{figure}
\center \epsfig{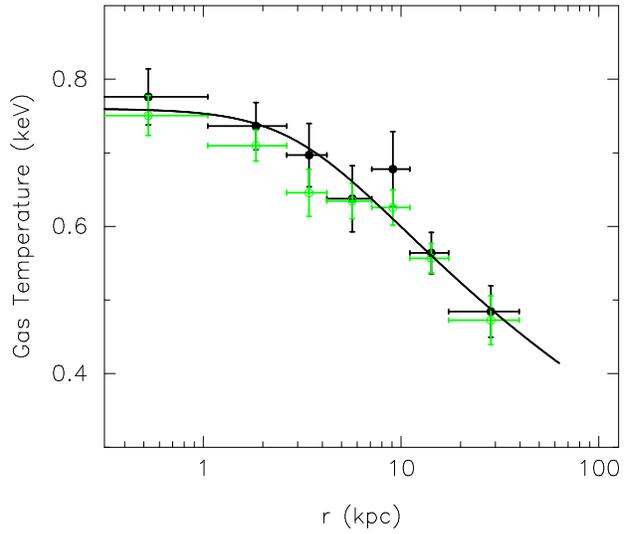}
\caption{Temperature profile of NGC\,6482.  Bold symbols show the
de-projected profile. The light symbols represent the projected
temperature profile. The curve represents the polytropic fit to the
de-projected temperature profile with $\gamma=1.14$.}

\label{fig7}
\end{figure}

\subsubsection{Temperature profile}

Fig. \ref{fig7} shows the annular and de-projected temperature
profiles out to 150 arcsec. The temperature profile shows a drop 
outward reaching 0.63 of its central value at a mean radius of 29~kpc. 
It is useful to have an analytical temperature profile, which can be
used to derive the entropy profile. We adopt a polytropic model:
\be
T(r)=T(0)[1+(\frac{r}{r_c})^2]^{-3/2\beta(\gamma-1)},
\ee
where $r_c$ and $\beta$ are the same parameters as one obtains from a
$\beta$-model to the surface brightness profile.  We fixed $\beta$ at 
an average value
of 0.53, from the inner and outer regions, and fitted the polytropic
temperature profile, equation 6, with the core radius and $\gamma$
exponent as free parameters.  We found $\gamma=1.14$ and $r_c=12.5$
arcsec. The profile is presented in Fig \ref{fig7}.

\subsubsection{Central annulus and the point source}
The innermost region gave a poor spectral fit. Here we investigate two
possible causes. The temperature profile shows no sign of cooler gas
in the central region, contrary to many X-ray groups of similar mass
\citep{hp00a,sun03,mush03}. 
Since the point source detection routine found a point source at
the centre of the galaxy in both low and high energy bands, we suspected
that it might contain an active galactic nucleus (AGN). There is further
motivation for this in the literature -- \citet{gou94} found an
$H\alpha+[NII]$  flux of $10.8(\pm1.8)\, \times 10^{-14}$ 
erg~cm$^{-2}$~s$^{-1}$ in the core of NGC\,6482.

To check whether a central point source could be contaminating our
spectral fit in the innermost region, which extends to $r=4$~arcsec,
we excluded the central 2~arcsec, which encompasses most of the point
spread function of \Chandra, and monitored the fit quality. We found no 
significant improvement in the fit compared to the original reduced \ki 
of 1.12. The limited number of counts from the central point source 
does not permit any further analysis of a possible AGN component. An 
upper limit for the X-ray emission from the point source is estimated
assuming a canonical AGN contribution (a power law spectral  model
with photon
index=1.7) in the spectrum of the innermost region. The spectrum is
consistent with a maximum AGN luminosity of $\approx 1.0
\times 10^{40}$ ergs/s within 90\% confidence interval 
(i.e. $\Delta$\ki=2.71).

A second possibility is that gas in the high density core
is cooling out. We therefore included a 
cooling flow component, and fitted the central annulus
with an absorbed APEC+CFLOW model. This fitted with 
$\dot{M}=3.21_{-2.06}^{+2.50}$ M$_\odot$ yr$^{-1}$ and a slightly
increased intrinsic column of $n_H=0.19\times10^{22}$~cm$^{-2}$, 
giving an improved reduced \ki of 1.03 with 30 degrees of freedom.
The implications of this are discussed later in the paper.

\begin{table}
\begin{center}
\caption{Results from the PROJCT spectral fit to the seven annuli. 
\label{table1}
}
\begin{tabular}{llccr}
\hline
Annulus & mean radius & temperature & $L_{X,bol}$& intrinsic $n_H$   \\
       & ($h_{70}$kpc) & (keV) & $10^{41}$ergs/s & $10^{22}$ cm$^{-2}$ \\
\hline
1   &  0.54  & $0.776\pm0.037$ &  0.64 & $0.177\pm0.069$ \\
2   &  1.88  & $0.736\pm0.032$ &  0.88 & $0.080\pm0.066$ \\
3   &  3.48  & $0.696\pm0.053$ &  0.62 & $0.202\pm0.073$ \\
4   &  5.76  & $0.637\pm0.045$ &  0.71 & $0.164\pm0.066$ \\
5   &  9.25  & $0.677\pm0.049$ &  0.65 & $0.119\pm0.075$ \\
6   &  14.47 & $0.563\pm0.029$ &  0.65 & $0.003\pm0.050$\\
7   &  28.94 & $0.481\pm0.035$ &  1.05 & $0.000\pm0.081$\\

\hline
\end{tabular}
\end{center}
\end{table}

\subsection{Entropy}

\begin{figure}
\center
\epsfig{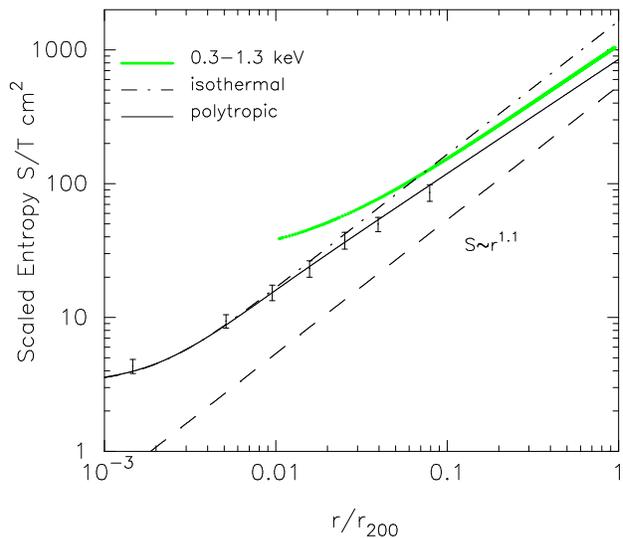}
\caption{Entropy profile from 7 successive annuli. Solid and dot-dashed 
lines are the entropy profiles assuming polytropic and isothermal
temperature profiles, respectively. The thick grey line is the
mean profile for a sample of cool (0.3-1.3~keV) systems derived
by \citet{psf03}, and dashed line represents the typical profiles
seen in cosmological simulations including only gravity and shock
heating.}

\label{fig8}
\end{figure}

The entropy of the X-ray emitting gas is defined here as:

\be
S(r)=kT(r)/n_e(r)^{2/3}\, {\rm keV cm}^2 ,
\ee
where $n_e$ is the electron density. The polytropic model
described above was used to obtain the temperature at a given
radius. Similarly the electron density was obtained from the the
$\beta$-model gas density profile. The resulting parameterised
entropy profile, scaled by $1/T$ (where $T=$0.66~keV is the mean
temperature) to facilitate comparison with other groups and clusters,
is shown in Fig \ref{fig8} as the solid line. 

For comparison with other systems, it is also useful to scale
such profiles to the virial radius, defined here as the radius,
$r_{200}$, within which the mean total density of the system is
200 times the critical density of the Universe at the present time.
In section 4.1 below, we will derive the value of $r_{200}$ from the
present Chandra data, by extrapolating fitted models for gas density
and temperature to derive a total mass profile. However, this
involves an extrapolation of these models by a factor of $\sim10$
in radius, beyond the extent of the Chandra spectral data which we fit.
We therefore prefer (except in section 4.1) to use the value
$r_{200}=361$~kpc, derived by \citet{sand03} from ROSAT spectral imaging
data extending to a radius of 100~kpc, which has been derived excluding a
large region at the centre, and this has been used to scale 
the x-axis in Fig~\ref{fig8}.

To estimate the average entropy in each shell directly from the
data, the density was obtained in the following way. 
The de-projected luminosity of each 3D shell was combined with the
model emissivity  $\epsilon$ at the shell temperature 
(and in the same energy band, 0.3-2.5 keV)
and the shell volume $V$, via:
\be
L=n_e n_H V \epsilon ,
\ee
to derive the electron density. This was then combined with the
deprojected temperature, to give a spot value for the entropy.
The resulting entropy values and their associated errors are 
presented in Fig~\ref{fig8}, and agree well with the analytical curve.
The dot-dash line shows the entropy profile assuming an
isothermal model, and clearly overestimates the entropy outside
the innermost regions, which is reduced by the steep decline
in temperature. 

Under the assumption of self-similarity of cluster properties,
$S(r/r_{200})$ profiles should scale as $1/T$, so we can compare
directly with scaled models from simulations of rich clusters.  Such
simulations, as well as analytical models of spherical accretion onto
clusters \citep{toznorm01}, predict entropy profiles with slope
$S\propto r^{1.1}$, outside a central region in which the profile may
flatten. The dashed line in Fig~\ref{fig8} shows such a profile,
normalised using simulations by Scott Kay (private communication) 
including only
gravitational processes and shock heating.  Clearly the entropy of the
gas in NGC\,6482 is substantially higher than would be expected from
such self-similar scaling, as seems to be the case in all groups
\citep{psf03}.

The mean $S/T$ profile derived for a set of cool (0.3-1.3 keV) systems
by \citet{psf03} is also shown in Fig~\ref{fig8}, and lies somewhat
above the profile seen in NGC\,6482. However, as has been pointed out
by \citet{sun03} and \citet{mush03}, a large amount of scatter is seen
in the entropy within galaxy groups.
Comparison with other entropy profiles of poor groups based on 
data from \XMM and \Chandra, is instructive. Converting where necessary
to $H_0=70$ km~s$^{-1}$~Mpc$^{-1}$, the result for $S$(0.1$r_{200}$) reported 
for NGC\,1550 ($T=1.37$~keV) by \citet{sun03} is 120~keV~cm$^2$, whilst
\citet{mush03} find values of 107~keV~cm$^2$ and 250~keV~cm$^2$ respectively
in NGC\,4325 ($T=0.95$~keV) and NGC\,2563 ($T=1.36$~keV). The entropy of 
NGC\,6482 at the same normalised radius is $89\pm12$ keV cm$^{2}$, which
is lower than any of the above. 

However NGC\,6482 is cooler than these other groups.
\citet{psf03} find that the entropy scaling law in groups and clusters
is well represented, on average, by $S\propto T^{0.65}$, rather than
the self-similar relation $S\propto T$. Applying a $1/T^{0.65}$
scaling to the four groups discussed above, we find that NGC\,6482 has
a scaled entropy well below that of NGC\,2563, but slightly higher than
NGC\,1550 and NGC\,4325.

A striking feature of the profile in Fig~\ref{fig8}, compared to the
entropy profiles seen in most simulations, or in the averaged 
observational profiles derived by \citet{psf03}, is the absence of 
any significant core to the entropy distribution. The entropy appears
to drop all the way into the centre, to the limit of the resolution of
our \Chandra data, corresponding to $\sim 10^{-3}$r$_{200}$.
Similar behaviour was reported in NGC\,1550 by \citet{sun03}, however, in
NGC\,1550 the temperature drops inside 0.05r$_{200}$, so that
the continuing entropy decline could be due to the effects of radiative
cooling, whilst in NGC\,6482 the temperature rises continously into the 
centre.

\begin{figure}
\center
\epsfig{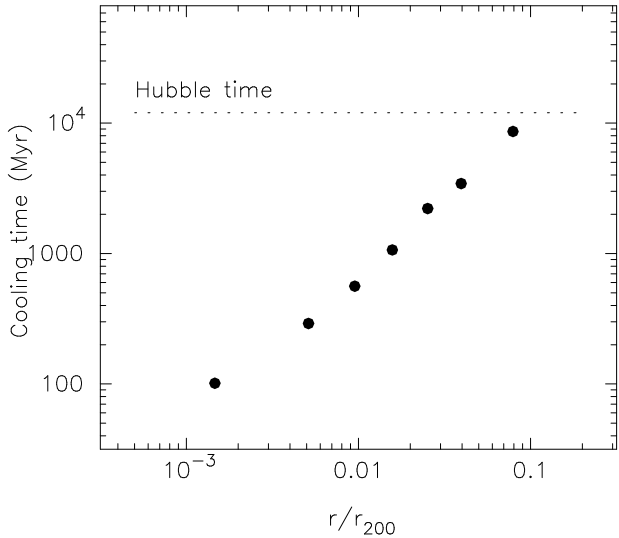}
\caption{Cooling time profile of the core. The dotted line represents 12 Gyr.}
\label{fig9}
\end{figure}

\subsection{Gas cooling time}
The cooling time of the gas was calculated by dividing the total
thermal energy of each shell by the associated de-projected
luminosity. The resulting profile is presented in Fig. \ref{fig9}.
The cooling time of the entire core, within 0.1$r_{200}$, is less than
a Hubble time, and it drops to $\sim 10^8$~years within the region
resolved by \Chandra.

\section{Gas and total mass profiles}

We use the gas density profile and temperature profile to derive the
gas and total gravitational mass of the system assuming hydrostatic
equilibrium and spherical symmetry. The total mass is given by:
\be
M_{grav}(<r)=-\frac{kT(r)r}{G\mu m_p}[\frac
{dln\rho(r)}{dlnr}+\frac{dlnT(r)}{dlnr}].
\ee
where G and $m_p$ are the gravitational constant and proton mass and
$\mu=0.6$.

It is straightforward to derive an analytical description of the mass
distribution using our analytical expressions for density and
temperature. However estimating the error in the mass for the data
points is not similarly simple, since there are contributions
from uncertainties in both temperature and density, and the constraints
of physical validity should ideally be taken into account. 

\citet{nb95} have suggested a Monte-Carlo method to
derive the empirical mass profile and the associated errors. The
gravitational mass consists of two parts. One is the variation in gas
density, $d\ln n/d\ln r$ and the other is that of the temperature, $d\ln
T/d\ln r$. For the latter we have generated 1000 physical temperature
profiles with a temperature window of 0.05 keV and step size of 8 arcsec
within the observed temperature range. A physical temperature is 
one which guarantees a monotonically increasing mass with radius.  In
this simulation the density profile parameters were fixed at the
values from the $\beta$-model fit.  In order to estimate the contribution 
of the density variation ($d\ln n/d\ln r$) to the mass error, the surface 
brightness profile at the location of each data point
was fitted with $d\ln n/d\ln r$ as the free parameter instead of
$\beta$. The error was estimated at the 68\% confidence level using
the error matrix provided by the fitting program. We checked the results
by monitoring the \ki variation. In this experiment the values of
other parameters were optimised. The total error in the mass at each
data point was then derived by adding the two uncertainties quadratically.

\begin{figure}
\center
\epsfig{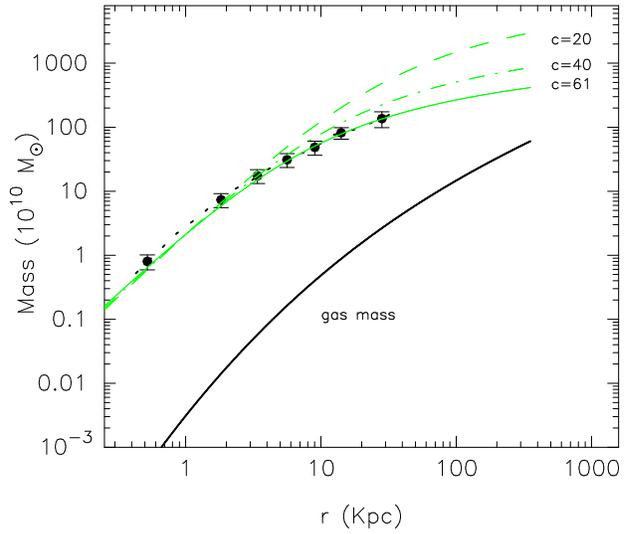}
\caption{The best fit NFW mass profile is shown by gray (green) solid
curve. The concentration parameter is $c\approx 60$. The dashed-dotted
and long dashed lines represent the NFW profiles with lower
concentrations 40 and 20, respectively. The tick solid curve shows the
total gas mass profile. All the profiles extend to 361 kpc.  The
errors in the mass were estimated using MC simulations. The dotted
line is the mean mass from the MC simulation.} 
\label{fig10}
\end{figure}

\subsection{NFW mass profile and concentration}

Motivated by the results of numerous cosmological N-body simulations, we
attempt to fit a NFW profile \citep{nfw95}
\be
\rho_m(r)=\frac{4\rho_m(r_s)}{(r/r_s)(1+r/r_s)^2},
\ee
to the total gravitational mass density.
We can then obtain the following integrated mass profile for a spherical mass 
distribution, 
\be
M_{tot}(<r)=16\pi\rho_m(r_s)r_s^2[r_s \ln (1+r/r_s)-\frac{r}{1+r/r_s}].
\ee
where $\rho_m(r_s)$ is the density at $r_s$. The mass concentration  
parameter can then be defined as $c=r_{200}/r_s$. By definition, $r_{200}$ is 
the radius at which the mean gravitational mass density is 200 times the 
critical density, $\rho_c(z)$. For our assumed cosmology 
($\Omega_m$=0.3, $\Lambda$=0), the critical density is 
$\rho_c(z)=\frac{3H_0^2}{8\pi G}(1+z)^2(1+z\Omega_m)$. 

We fit the NFW mass profile, equation 11, to the data presented in Fig
\ref{fig10} to obtain $r_s$ and derive the the mean density profile.
The best fit profile has $r_s=5.1$ kpc, and is shown in Fig
\ref{fig10} as a thin solid line. $r_{200}$ is then calculated by
extrapolating the profile to 200 $\rho_c$. This gives $r_{200}\approx
310$ kpc, which is consistent within errors with the 
value of $r_{200}=361^{+181}_{-105}$ kpc derived from ROSAT data 
by \citet{sand03}, and used elsewhere in the present paper.
The extrapolated total mass at $r_{200}$ is  $M_{200}\approx 
4\times10^{12}$ M$_{\odot}$.

This best fit profile therefore has a concentration parameter
$c=310/5.1=61$, which is remarkably high. For comparison, Fig \ref{fig10}
also shows NFW mass profiles with concentration parameters of 40 and
20. These are clearly inconsistent with our data.

\subsection{Gas fraction}

The integrated gas mass can be simply derived by integrating the
gas density (equation 2), and the result is shown in Fig \ref{fig10}.
The most reliable measure of the gas mass fraction is that within a
radius of $0.1r_{200}$, since the temperature profile and
hence total gravitational mass is limited to this radius. We found
$f_{gas}(0.1r_{200})=0.02$. A similar gas fraction is seen in 
NGC\,1550 \citep{sun03} and NGC\,4325, but is significantly higher than
that found in  NGC\,2563 \citep{mush03}. Extrapolating implies a gas
fraction which rises to 0.05 at 0.3$r_{200}$, approximately 
double the typical value derived at this radius for cool 
systems by \citet{sand03}.

If we make a major extrapolation (factor of five for our surface
brightness data, which primarily determines the gas density,
and factor 10 for our spectral data, which are required to
constrain total mass) of our models to $r_{200}$, we derive a total
gas mass $M_{gas}(<r_{200})=6.7\times 10^{11} M_{\odot}$, and
infer a gas fraction $f_{gas}(r_{200})=0.16$. The latter is at least
as large as that seen in rich clusters, and certainly considerably
higher than that normally reported for galaxy groups.

\begin{figure}
\center
\epsfig{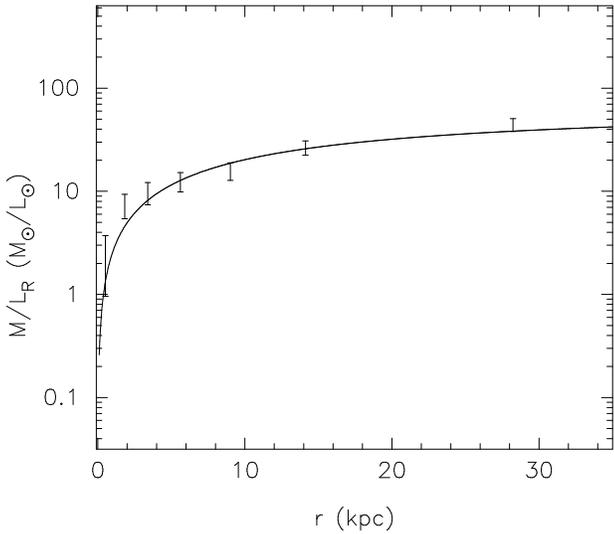}
\caption{Cumulative mass-to-light ratio profile. The errors are
estimated from the errors in mass and luminosity for each shell. The
solid curve was obtained by dividing NFW profile by the luminosity profile 
of the brightest group galaxy.}
\label{fig11}
\end{figure}

\subsection{Mass to light ratio profile}

With the high spatial resolution of the Chandra data, it is possible to
derive the mass-to-light ratio profile in the core of the
system. We analysed archived R-band CCD imaging data from the 0.9m JKT
to model the cumulative luminosity profile of the brightest group
galaxy. The photometric accuracy, based on calibration using five nearby 
stars in the field, was $\approx$0.1 mag. 

Fig \ref{fig11} shows the cumulative mass-to-light ratio
profile within $0.1r_{200}$. To obtain a continuous profile 
(the solid curve), we divided the fitted NFW mass profile by a numerical model 
fitted to the R-band luminosity of the galaxy. 

The profile shows a rapid rise in mass-to-light ratio with radius in
the central 5 kpc. Taking into account the luminosity of the other
known group members, the mass-to-light ratio in the R band at
$r_{200}$ rises to $71\pm15$~$M_{\odot}/L_{\odot}$.  This is
equivalent to 86$\pm$15~$M_{\odot}/L_{\odot}$
in the B band, assuming colours of a typical
elliptical. This value is within the scatter of the mass-to-light
ratios found by \citet{sp03} for groups and clusters,
and is significantly smaller than the R-band mass-to-light ratios of
250-450 found by \citet{vikh99} for similar systems.  The
\citet{vikh99} mass measurements were based partly on temperature
estimates from an $L_X-T$ relation, and are thus less reliable than
our mass measurements based on an accurate temperature profile.

\section{Scaling relations}

The X-ray scaling relations of groups differ from those of clusters.
Here we investigate whether this fossil group follows the scaling
relations of groups, or whether it supports the suggestion of \citet{jones03},
that there are significant differences between fossil
groups and normal groups.

\subsection{X-ray luminosity}

From the spectral analysis we find the total X-ray luminosity,
0.5-8.0 keV, to be $5.21\times10^{41}$ ergs/sec within
$\sim\!0.1r_{200}$. Beyond this the total X-ray luminosity of the
group has been estimated by extrapolating the surface brightness
profile from the $\beta$-model fit to $r_{200}$. 
We find $L_{X,bol}=1.11\times10^{42}$ ergs~s$^{-1}$ within $r_{200}$.

\subsection{The $L_X-L_{opt}$ relation}

The brightest group galaxy has a total R-band luminosity of
$4.46\times10^{10}L_{\odot}$, based on our analysis of the archived
JKT observation. This means that NGC6482 follows the same trend as the
six other fossil groups studied in
\citet{jones03}, in having an X-ray luminosity significantly higher than
normal groups of the same optical luminosity.

The high X-ray luminosity in NGC 6482 is clearly due to the high gas
density and is not a result of a high mass-to-light ratio.  The
central density of 0.15$\pm$0.06 cm$^{-3}$ is higher than that in any
group or cluster in the compilation of \citet{sand03}. 

\subsection{The $L_X-T$ relation}

\citet{jones03} found that two of their fossil groups were overluminous
compared to the expectations from the cluster $L_X-T$ relation, in
contrast to the majority of groups which may be underluminous compared
to the cluster relation. With an overall temperature of 0.7 keV 
the NGC\,6482 system falls within the scatter of previous measurements 
of the group $L_X$-$T$ relation \citep{mz98,
hp00b} and is not significantly offset.
Thus not all fossil groups are offset from the $L_X$-$T$ relation.
A possible explanation may be that the  NGC\,6482 system was 
essentially optically selected, whereas the \citet{jones03}
systems were X-ray selected, and their study may therefore have 
preferentially selected high $L_X$ fossil groups.

\section{Discussion and Conclusions}

In this section we first give a summary of the results from this 
detailed study of the nearest fossil group using high resolution 
Chandra data and discuss some of them in detail:

\itemize

\item The system has extended X-ray emission similar to bright X-ray 
groups but the optical light is dominated by the central giant 
elliptical galaxy. 
 
\item  The uniform and circularly symmetric distribution of the hardness 
ratio suggests that the system is relatively relaxed with no sign of 
recent violent perturbation.

\item This is the first Chandra resolution observation of a relaxed system in 
which the temperature decreases continually outwards despite a central 
cooling time of only $\sim 10^8$ years.  

\item In agreement with recent studies of low temperature systems, the entropy 
profile shows no flat core.  

\item The Mass-to-light ratio is $71\pm15$ ~$M_{\odot}/L_{\odot}$ (R-band) at $r_{200}$.

\item The mass concentration parameter has a remarkably high value of $c\approx 60$, suggesting an early epoch of formation.

The proximity of NGC\,6482, coupled with the high angular resolution
of \Chandra, has allowed us to resolve the X-ray structure of the system
in a way never previously achieved with any fossil group. We find
that the high value of $L_X/L_{opt}$, in this fossil group at
least, results from a very high central gas density (0.15 cm$^{-3}$),
whilst the mass-to-light ratio, measured accurately here, is not exceptional.

The internal absorption may be due to the substantial gas and dust
inferred from optical data \citep{gou94}. We estimate a hydrogen 
column density of $\approx1.7\times10^{20}$ cm$^{-2}$ based on the 
optical extinction,
well below the absorption derived from our X-ray spectra. A plausible
explanation may be the presence of a diffuse `hot dust' component in
the central few kpc, as suggested by IRAS \& ISO data in many
ellipticals \citep{gou95,ferr02}. The small size dust grains of such a
component would be undetected using optical extinction measurements.

The centrally-concentrated gas density profile, coupled with a $T(r)$
profile which rises continuously into the centre of the group, implies
a remarkably centrally-concentrated total density profile, with an NFW
concentration parameter $c\approx 60$. This suggests that the system
was formed at an early epoch, has been relatively undisturbed since,
and thus reflects the high density conditions of the early Universe.
Although N-body simulations involving cold dark matter predict that
low mass systems such as NGC\,6482 (we measure a total mass of about
$4\times10^{12}$ M$_{\odot}$) should form before rich clusters, and
hence have higher concentration parameters, the properties of  NGC\,6482
seem altogether exceptional. For example, \citet{wechs02} identified
a total of 14,000 halos in a cosmological simulation, and studied 
their properties. They found strong correlations between halo mass,
formation epoch and concentration parameter, in the sense that lower
mass halos tend to form earlier, and have higher $c$. However, the highest
values of $c$ for halos from their simulation, in the mass range 
3-$4\times10^{12}$ M$_{\odot}$, was $c\sim 25$. Such halos formed
(according to their definition of `formation time', based on the rate
of fractional mass growth) at a redshift $z\sim$5.

How are we to understand the extremely high concentration of NGC\,6482?
It is possible that this is pointing to some deficiency in
cosmological simulations. Some previous authors have reported
surprisingly high values of $c$, compared to those expected from
simulations. For example \citet{david01} found $c\approx12$ in the
moderately rich cluster A780, which is three times greater than the
expected value assuming a $\Lambda CDM$ cosmology, and 
would require an abnormally early redshift of formation of $z\approx 4$. 
\citet{parnud02} found a mass 
concentration of $c\approx 5.4$ in their study of A1413 by 
fitting an NFW profile. Similarly they obtained $c\approx 3.75$ 
for A1983 \citep{parnud03}.
In the present case, at least, it is worth reflecting
that the scale radius of 5.1~kpc lies well within the central galaxy, 
in a region where (c.f. Fig.~\ref{fig11}) the stellar mass is the dominant
component. Hence the scale radius, and the concentration parameter
calculated from it, may be significantly affected by dissipative processes
taking place in the baryonic component of the system. Such processes
are known to be poorly modelled in cosmological simulations, where
they are included at all. For example, it is quite conceivable that 
inflow of gas into the centre of NGC\,6482, such as is believed
to take place during galaxy mergers, could deposit large quantities
of cooling gas in the inner regions of the galaxy, triggering a nuclear
starburst and steepening the central mass density profile.
Such processes may well have raised the value of $c$ subtantially,
however the $D_{25}$ radius of the galaxy is only $\approx16$ kpc, so we
can have some confidence that the primordial scale radius must have
been no larger than this, and therefore that the concentration
parameter must have been high ($c\geq$20), indicating an early formation
epoch for the system.

Thus the NGC\,6482 system is an old group, giving sufficient time 
for dynamical friction to act on the most massive galaxies within the
group, causing them to migrate to the centre, and merge to produce
the giant elliptical galaxy NGC\,6482, leaving no other luminous galaxies
within 2 mag of the dominant galaxy.

As we have seen, the entropy of NGC\,6482 at $r=0.1r_{200}$ appears to
be lower than the average for groups, but not outstandingly so,
considering its low mean temperature. The most interesting feature is
the way in which the entropy drops all the way into the central 1/2
kpc of the galaxy, where our observations become limited
by the spatial resolution
of \Chandra.  This conflicts directly with the expectations of simple
preheating models \citep{evrard91, kaiser91, cavaliere97, balogh99,
valageas99, toznorm01} in which gas is placed on a high adiabat 
before the formation of the system, setting a lower limit (argued by 
\citet{PCN99} to be $\sim100$~keV~cm$^2$) below which its entropy cannot 
drop, in the absence of significant cooling. Similar results have been 
reported recently by a number of authors \citep{psf03,parnud03,sun03,
mush03}, however, in the case of NGC\,6482, the behaviour is all the 
more striking, given that the temperature of the IGM rises continuously
inward, so that the low central entropy cannot be attributed to the
effects of cooling. It appears that simple global preheating models 
are no longer tenable, though this does not prohibit models in which
local heating, either inside or outside forming groups and clusters,
is responsible for breaking the similarity scaling between galaxy
systems of different mass. Even NGC\,6482 has an entropy profile
which lies well above that expected from self-similar scaling of clusters.

Despite the fact that the entire core shows a cooling time less than
the Hubble time, and in the innermost regions resolved by \Chandra it
drops to $\sim 10^8$~years, there is no evidence for any cooler gas at
the centre of the galaxy. As a result of recent studies with \XMM
and \Chandra \citep{peterson01,boh02,sak02,peterson03} it is now 
well-established that gas in the cores
of most clusters does not appear to be cooling at the rates often
inferred in the past. The reasons for this are still a topic
of lively debate \citep{fab01}. Despite the general lack of really
cool gas, a decline by up to a factor $\sim 2$-3 in temperature
is commonly seen in the cores of clusters and groups. The lack of
any decline at all in NGC\,6482, immediately eliminates one
suggested mechanism for counteracting central cooling. Thermal
conduction \citep{voigt02} cannot act to transport energy
into the core against the temperature gradient, and so is ruled
out. A highly inhomogeneous metallicity distribution \citep{morris03}
can lead to a lack of lines from cool gas phases, and hence limit
the apparent temperature drop in the central cooling regions,
but it cannot suppress the signature of cooler gas altogether.
Heating by cluster merging \citep{kempner03,gomez02} can disrupt 
cooling gas, and 
inject energy which may suppress cooling for some time \citep{kp97}, 
however it seems most unlikely that this could
explain the situation seen in such an apparently old, relaxed system
as NGC\,6482, given the symmetric hardness ratio map.

Three sources of heating remain, which might help to offset the
observed radiative energy losses ($\sim 3.5\times 10^{41}$~erg~s$^{-1}$
from within the central 10~kpc).  These are PdV work, 
supernovae and AGN.
Given the highly peaked nature of the total density profile, gas will
be subject to considerable $P\,dV$ work as it flows inward through the
core of the system. To explore the effects of this, we seek
a steady state cooling flow solution by
solving the energy conservation equation \citep{fab84}, 
\begin{equation}
\rho v \frac{d}{dr}(H(r)+\phi(r))=n_e(r)n_H(r)\epsilon(T) ,
\end{equation}
where $H(r)=\frac{5}{2}\frac{kT(r)}{\mu m_H}$ is the specific enthalpy and 
$\phi$ is the gravitational potential, coupled with mass conservation 
\begin{equation}
\dot{M}=4\pi\rho v r^2={\rm constant}
\end{equation}
for the observed gas density (equation 2) and the gravitational potential
inferred from our data, to calculate the temperature profile. 

\begin{figure}
\center
\epsfig{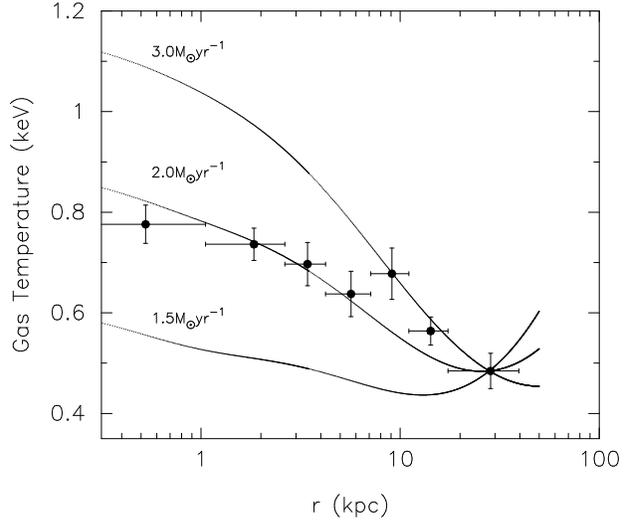}
\caption{Model temperature using the observed mass and gas density 
profiles for different mass accretion rates. The data point are the 
de-projected temperatures in Fig \ref{fig7}.}
\label{fig12}
\end{figure}

We fix the gas properties at the outer edge of our analysis
(where the cooling time is approximately equal to the Hubble time)
to match our observations, and seek for a value of $\dot{M}$ which
produces a $T(r)$ profile matching that
which we observe (Fig. \ref{fig12}), using the integral equation
\begin{equation}
kT(r)=\frac{2\mu m_H}{5}[\phi(r)-\frac{4\pi}{\dot{M}}\int_{r_7}^{r} 
1.17 n_H^2(x)\epsilon(T) x^2 dx] + C ,
\end{equation} 
where $r_7=28.94$ kpc is the mean radius of the outermost shell and the constant 
of integration C is chosen to normalise the temperature profiles to the 
value observed at $r=r_7$. 
A good match to our profile is obtained (Fig.~\ref{fig12}) with
$\dot{M}\approx2$ M$_{\odot}$ yr$^{-1}$. Lower mass accretion rates, 
$\dot{M}\approx 1.5$, result in a nearly isothermal core, whilst
higher rates give too steep a profile. Under this steady state
cooling flow model, the gas is heated by the gravitational
work done on it as it flows inward, before cooling and dropping
out at the centre of the galaxy. The spectral results shown discussed
above in Section~3.2.2, do not disallow such a cooling component
in the centre of the system -- in fact, as we have seen, it actually
improves the spectral fits. 

Steady-state cooling flow models have fallen out of favour in recent
years, as discussed above, and it is, of course, possible that some
processes other than $P\,dV$ work are responsible for maintaining
the steep temperature profile. Any steady-state solution other than
that which we have derived above would have to involve distributed
heating, since the gas must be prevented from cooling throughout
the region we have analysed.   

An obvious source of heating is supernovae (SN), which could be of
type I and/or II.  To estimate the contribution of SNIa, we use a
total SN I rate of ($0.18 \pm 0.05$)$h_{75}^2$ per century per
$10^{10}~L_{B\odot}$ and an energy production of $10^{51}$ erg per
supernovae \citep{capp99}. This corresponds to
$L_{SN}/L_{B}=2.5\times10^{30}$ erg s$^{-1}$ L$_{\odot}^{-1}$ after
correcting for the value of $H_0$ used in this study, giving an SNIa
contribution of $\sim 1.0\times 10^{41}$ erg s$^{-1}$. Hence SNIa
could contribute a significant fraction of the 
$\sim 3\times 10^{41}$~erg~s$^{-1}$ radiated by the region under study.
 
Using the $H\alpha$ luminosity \citep{gou94} and associated star 
formation rate, SFR(M$_\odot$/yr$^{-1}$)=$7.9\times10^{-42}$ L$_{H\alpha}$
(erg s$^{-1}$) \citep{cardiel03}, we estimate the SNII contribution assuming a 
SN rate $9.26\times 10^{-3}$ times the star formation rate \citep{kawata01},
and an energy of $10^{51}$ erg per SN.  This corresponds to a 
heat injection of $0.14\times 10^{41}$ erg $s^{-1}$, which is 
much smaller than the power required to replace the radiative losses.

The final source of heating which might play a role is injection of energy
from an AGN. AGN heating is achievable either by Compton heating,
mechanical heating or viscous dissipation of pressure waves generated by
AGN (Ruszkowski \etal 2003). \citet{binney95} have argued that the
last heating mechanism is most efficient. These authors, and several
others, have shown that even a small fraction of AGN mechanical energy
transferred to the ISM is sufficient to balance the radiative loss of
energy via X-ray radiation. The bubbles and jets released from AGN are
responsible for distributed heating, but it is important to note
that in reality such processes take place in a non-steady fashion.
 
A key feature of our observations is the negative temperature gradient
with radius. Heated cooling flow models are able to reproduce such a
negative temperature gradient (e.g. \citet{fbwm02}, but a
detailed comparison is required before more definitive conclusions can be
drawn. Thus a non-steady state solution remains a possibility. However for
a steady state solution, we have shown that PdV work alone could
generate a temperature profile similar to that observed in NGC\,6482.

\section*{Acknowledgments}
We would like to thank Ben Maughan and Andy Read for their advice
during the data analysis, and Mark Voit and Scott Kay for results and
insights relating to the scaling of entropy profiles from simulations.
We are also grateful to Ewan O'Sullivan and Alastair Sanderson for
their contributions to the early stages of this project. The authors
thank the anonymous referee for suggestions that improved the
presentation of the paper. We thank the \Chandra observatory team.
This research has made use of the NASA/IPAC
Extragalactic Database (NED) which is operated by the Jet Propulsion
Laboratory, Caltech, under contract with the National Aeronautics and
Space Administration. The optical images have been retrieved from the
Digitized Sky Survey.

\bsp

\label{lastpage}
\end{document}